\documentstyle[12pt,psfig]{article}

\renewcommand{\baselinestretch}{1.1}

\begin{document}

{\sc\noindent Helsinki Institute of Physics,\hfill
Preprint series,\hfill
HIP-1997-09, \hfill 6 March 1997}

\vspace*{1.5cm}

{\Large\bf\noindent Weisskopf-Wigner model for wave packet excitation}

\vspace*{1.0cm}

{\large\noindent Asta Paloviita$^{\dag\ddag}$, 
Kalle-Antti Suominen$^{\S}$ and Stig Stenholm$^{\dag}$}

\bigskip

{\noindent $^{\dag}$Helsinki Institute of Physics,
PL 9, FIN-00014 Helsingin yliopisto, Finland\\
$^{\ddag}$Present address: Nokia Research Centre, PL 100, FIN-33721 Tampere,
Finland\\
$^{\S}$Theoretical Physics Division, Department of Physics, University of
Helsinki, PL 9,\\ FIN-00014 Helsingin yliopisto, Finland} 

\vspace*{0.5cm}

\begin{center}

{\bf Abstract}

\end{center}

\begin{quote}
We consider a laser induced molecular excitation process as a decay of a
single energy state into a continuum. The analytic results based on
Weisskopf-Wigner approach and perturbation calculations are compared with
numerical wave packet results. We find that the decay model describes the
excitation process well within the expected parameter region.
\end{quote}

\section{Introduction}\label{sec:intro}

It has been known for a long time that when an isolated bound state is
interacting with a continuum of quantum states, the occupation of the bound
state experiences exponential decay; the first proper treatment was
given by Weisskopf and Wigner in 1930~\cite{r1}. A simple example of such a
situation is a two-state atom coupled to the electromagnetic modes of the
vacuum, which leads to spontaneous emission; see e.g.~refs.~\cite{Loudon,Stig}.
The general model has been described in detail within the formalism of
scattering theory in ref.~\cite{r2}, but this approach provides only a steady
state treatment. In a recent paper~\cite{r3} we considered this problem in a
genuine time dependent setting and discussed the decay process as a method to
prepare a moving wave packet on a molecular electronic state in accordance with
our earlier treatment of coupled channel molecular wave packet dynamics; for a
review see ref.~\cite{r4}. Now we proceed to study the validity of this
exponential decay description of laser induced wave packet excitation in
molecules.

\begin{figure}[tbh]
\vspace*{-1.0cm}
\centerline{\psfig{angle=90,width=6.0in,file=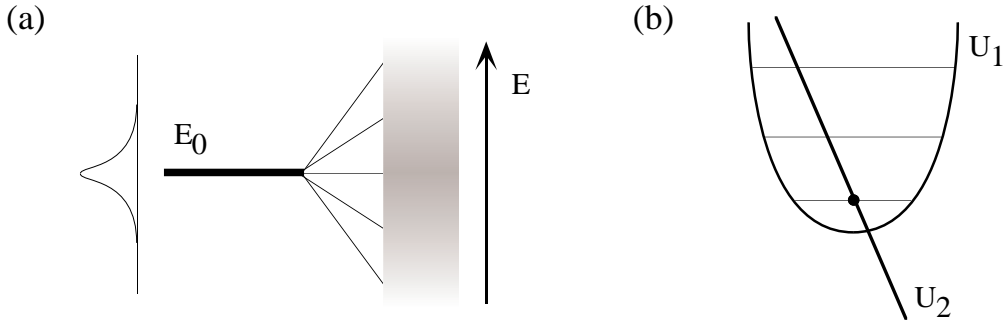}}
\vspace*{-1.0cm}
\caption{(a) An isolated energy level at $E_0$ is coupled to a system with a
continuous energy spectrum $E$. (b) In the molecular excitation model the
isolated state is the ground state of the molecule, $U_1$, and the continuum is
provided by the linear excited state potential $U_2$.
\label{schemes}}
\end{figure}

The prototype model for the Weisskopf-Wigner decay is shown in
fig.~\ref{schemes}(a); the discrete state embedded in a continuum acquires a
Lorentzian line shape, which corresponds to exponential decay of its occupation.
The theoretical description leading to exponential decay is based on an
approximation which represents the continuum spectral density by a single
dominating pole~\cite{r5,r6}. In the general case this is only an
approximation, which is expected to hold in the {\it perturbative limit} of weak
coupling between the discrete state and the continuum, in which case the result
of time dependent perturbation theory agrees with the Weisskopf-Wigner result.
When we go beyond this limit, the result is nonexponential decay~\cite{r2} or a
continuous transition of the decay into a periodic oscillational behaviour of
the Rabi type as discussed in ref.~\cite{r7}. Thus we need to establish for
molecular excitation the parameter regions where a) the perturbation approach
is applicable, and b) where the excitation process can be described as
exponential decay into the continuum.

In this paper we develop further the decay model introduced in ref.~\cite{r3}
and featured in fig.~\ref{schemes}(b). The lowest vibrational level of
the molecular electronic ground state corresponds to the discrete state; it
can be described adequately by the harmonic potential $U_1$. The molecular
counterpart of the continuum is an excited electronic state which can be
adequately described by the linear potential $U_2$ near the molecular
equilibrium position. Usually state 2 is a dissociating one, but it could
also represent a bound state in a energy region where the vibrational states
have a very small separation (quasicontinuum). These potentials are a good
basis for the molecular processes within the Born-Oppenheimer
approximation. If we also apply the dipole and rotating wave approximations to
the laser-molecule interaction, we can shift the excited state potential down
by one laser photon, which then leads to the elimination of the rapidly
oscillating field term from the coupling between the states 1 and 2; for more
details see ref.~\cite{r4}.

If the parameters of our model are chosen such that the vibrational excited
states of the harmonic potential remain mostly unoccupied during the
interaction, the decay of the oscillator ground state can be taken to be a
model of the level scheme in fig.~\ref{schemes}. The couplings between the
oscillator state and the eigenstates of the slope are not constant; the Condon
factors describing the overlap between the ground and excited state wave
functions have an energy dependence which complicates the model. Thus for
simplicity (one parameter less to consider) we arrange the sloping potential
$U_2$ so that a Franck-Condon transition at the origin of the coordinate system
gives the maximum overlap between the wave functions. Then we expect the Condon
factor to depend only slowly on energy around $E_0$ and the model will simulate
the behaviour of the Weisskopf-Wigner situation.

This paper is organised as follows. Section~\ref{sec:model} presents the basic
model. In sec.~\ref{sec:numerics} we give the results of our numerical
computations and discuss the occurrence of exponential decay. It is found that
this can be observed in a parameter range where the decay rate is less than a
definite value. This indicates that the exponential behaviour can indeed be
achieved only in a perturbative regime as expected. We also find that the
occupation of the state 2 emerges in the form of a localized wave packet even
beyond this perturbative parameter regime. In sec.~\ref{sec:analytic} we
discuss the analytic result for the decay and compare this with our results
obtained from numerical integration of our wave packet model. Finally we
summarize and discuss our results in sec.~\ref{sec:concl}.

\section{The model}\label{sec:model}

We consider a one dimensional system, spatial coordinate $x$, with two internal
states $\{|1\rangle, |2\rangle \}$. The Hamiltonian of the system is
written as 
\begin{equation}
   H=\left[ T_x+U_1(x)\right]|1\rangle \langle 1| +\left[
   T_x+U_2(x)\right] | 2\rangle \langle 2| +V\left[ | 2\rangle \langle
   1| +| 1\rangle \langle 2|\right] ,  \label{a1}
\end{equation}
where the coupling $V$ is chosen real. We define (in suitably chosen
dimensionless units, cf.~refs.~\cite{r3,r4}) 
\begin{equation}
   \begin{array}{lll}
      T_x & = & -\frac{\textstyle\partial^2}{\textstyle\partial x^2}, \\ 
      &  &  \\ 
      U_1(x) & = & \frac{\textstyle 1}{\textstyle 2}x^2, \\ 
      &  &  \\ 
      U_2(x) & = & \frac{\textstyle 1}{\textstyle\sqrt{2}}-\alpha x.
   \end{array}\label{a2}
\end{equation}
The Hamiltonian~(\ref{a1}) describes our chosen model as shown in
Fig.~\ref{schemes}(b). Since in our scaling the energy of the lowest
level of the harmonic potential is
\begin{equation}
   E_0=\frac 12\omega =\frac {1}{\sqrt{2}},  \label{a3}
\end{equation}
the energy of the linear potential resonates with the ground state at $x=0$,
see fig.~\ref{schemes}.

The eigenfunction of the ground state of the harmonic oscillator is 
\begin{equation}
   \varphi_0(x)=\frac{1}{\left( 2\pi^2\right)^{1/8}}\exp \left(
   -x^2/2^{3/2}\right)  \label{a4}
\end{equation}
with the energy eigenvalue (\ref{a3}). In the domain we are investigating, the
higher eigenfunctions are not supposed to play any role, and they are
consequently not needed, as already mentioned in sec.~\ref{sec:intro}.

The eigenvalues of the linear state $U_2$ form a continuum $\{-\infty
,\infty \}$ with the corresponding eigenfunctions easily obtained by a
Fourier transform 
\begin{equation}
   \tilde{\psi}_E(k)\equiv \int dx\,e^{\textstyle -ikx}\psi_E(x).  \label{a5}
\end{equation}
The functional form is that of the Airy function, and a suitable
normalization is given by 
\begin{equation}
   \psi_E(x)=\int \frac{dk}{2\pi\sqrt{\alpha}}\exp \left[ i\left(x+\frac{{\cal
   E}}\alpha \right)k-i\frac{k^3}\alpha \right] ,  \label{a6}
\end{equation}
where the energy parameter is 
\begin{equation}
   {\cal E}=E-\frac 1{\sqrt{2}}.  \label{a7}
\end{equation}
It is easy to prove the energy
normalization 
\begin{equation}
   \int \psi _{E^{\prime }}^{*}(x)\psi _E(x)\,dx=\delta (E-E^{\prime }).
   \label{a8}
\end{equation}
and the completeness 
\begin{equation}
   \int \psi _E^{*}(x^{\prime })\psi _E(x)\,dE=\delta (x-x^{\prime }).
   \label{a9}
\end{equation}
With these relations we see that the density of states is unity.

We assume that the system is prepared in the initial state 
\begin{equation}
   | i\rangle =\varphi_0(x)| 1\rangle ,  \label{a9a}
\end{equation}
which is then coupled to the continuum by the parameter $V$. Using the result of
the time dependent perturbation theory (Fermi Golden rule) the leakage into
state $| 2\rangle$, i.e. into our final state $|f\rangle$, is expected to
occur at the rate 
\begin{equation}
   \Gamma =2\pi \left| \langle i| H| f\rangle \right|^2 = 2\pi V^2\left|
   \langle \varphi_0 | \psi_{E_0}\rangle \right|^2,  \label{a10}
\end{equation}
where the continuum function is evaluated at the energy~(\ref{a3}) and the
parameter dependence is almost entirely in the Condon factor $\langle
\varphi_0 | \psi_{E_0}\rangle$.

From the normalization of eq.~(\ref{a6}) we expect the decay rate to scale as 
\begin{equation}
   \Gamma \propto 2\pi \frac{V^2}\alpha \equiv\Gamma_0,  \label{a11}
\end{equation}
where we have defined an {\it effective} decay parameter $\Gamma_0$.
When $\Gamma_0$ is below some limiting value, we expect that eq.~(\ref{a10})
gives a reasonable approximation to the decay rate of the initial
state~(\ref{a9a}). The result is modified by the Condon factor which will be
discussed below.

\section{Numerical calculations}\label{sec:numerics}

We have used the numerical approach described in sec.~2.3.3 of ref.~\cite{r4} to
propagate the initial state wave packet~(\ref{a9a}) on the coupled energy
surfaces of eq.~(\ref{a1}). However, instead of switching the coupling on
suddenly we have used the coupling term
\begin{equation} 
   V(t) = \left\{ 
   \begin{array}{ll}
      V_0\,{\rm sech\,}[(t-t_0)/T)],&t\leq t_0\\ &\\
      V_0&t>t_0
   \end{array}\right.\label{coupling} 
\end{equation}
where $t_0$ indicates in practice the beginning of the excitation process,
and we have set $T=0.05$ in our scaled units.

\begin{figure}[tbh]
\centerline{\psfig{width=3.4in,file=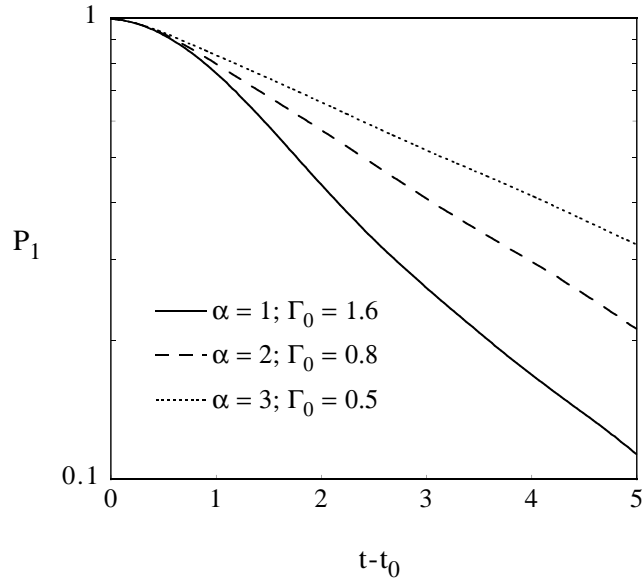}}
\caption[tat]{The occupation $P_1(t)$ for $V_0=0.5$. Note the
logarithmic scale on the vertical axis.}\label{v05}
\end{figure}

\begin{figure}[tbh]
\centerline{\psfig{width=3.4in,file=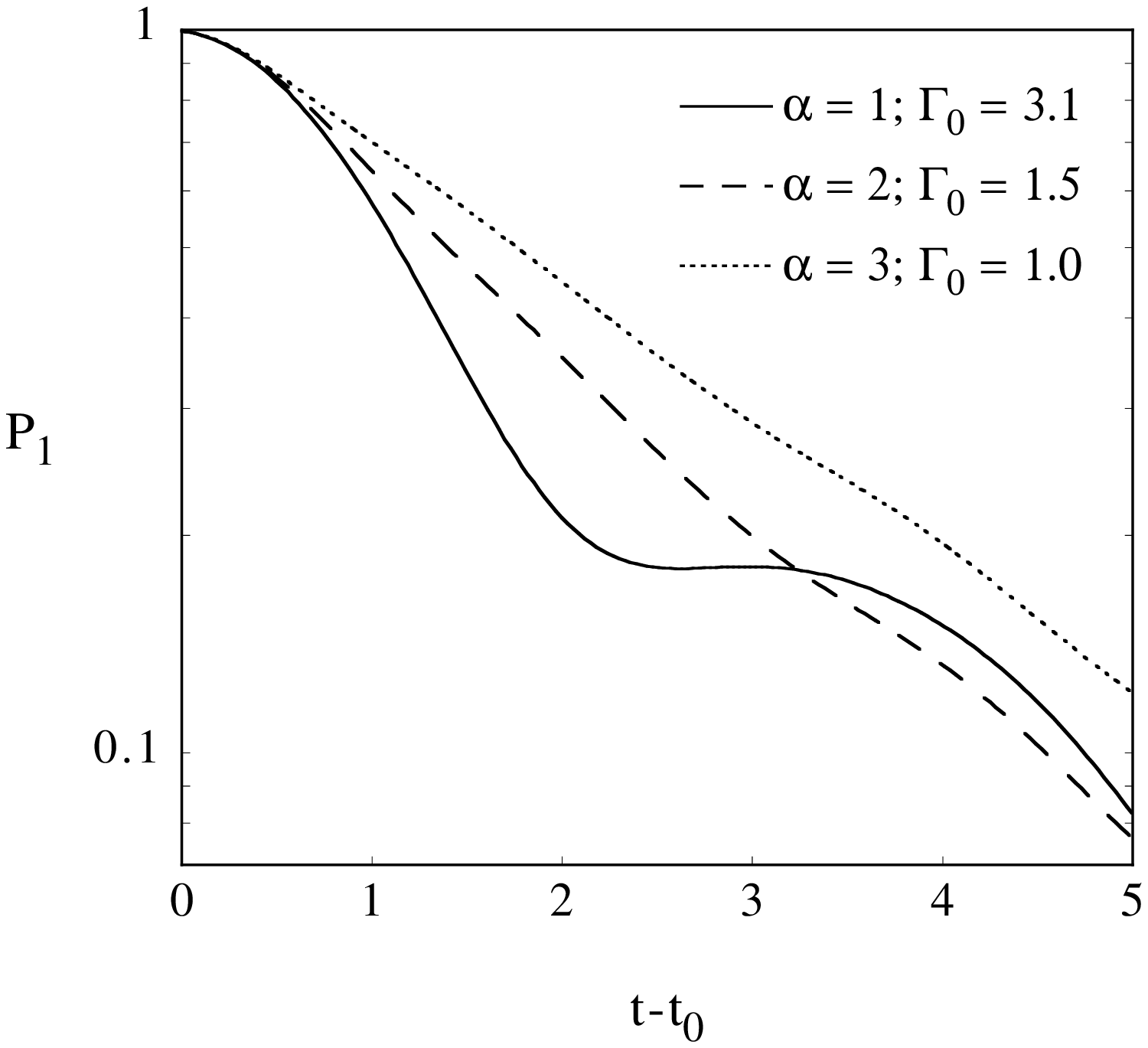}}
\caption[tut]{The occupation $P_1(t)$ for $V_0=0.7$. Note the
logarithmic scale on the vertical axis.}\label{v07}
\end{figure}

\begin{figure}[tbh]
\centerline{\psfig{width=3.4in,file=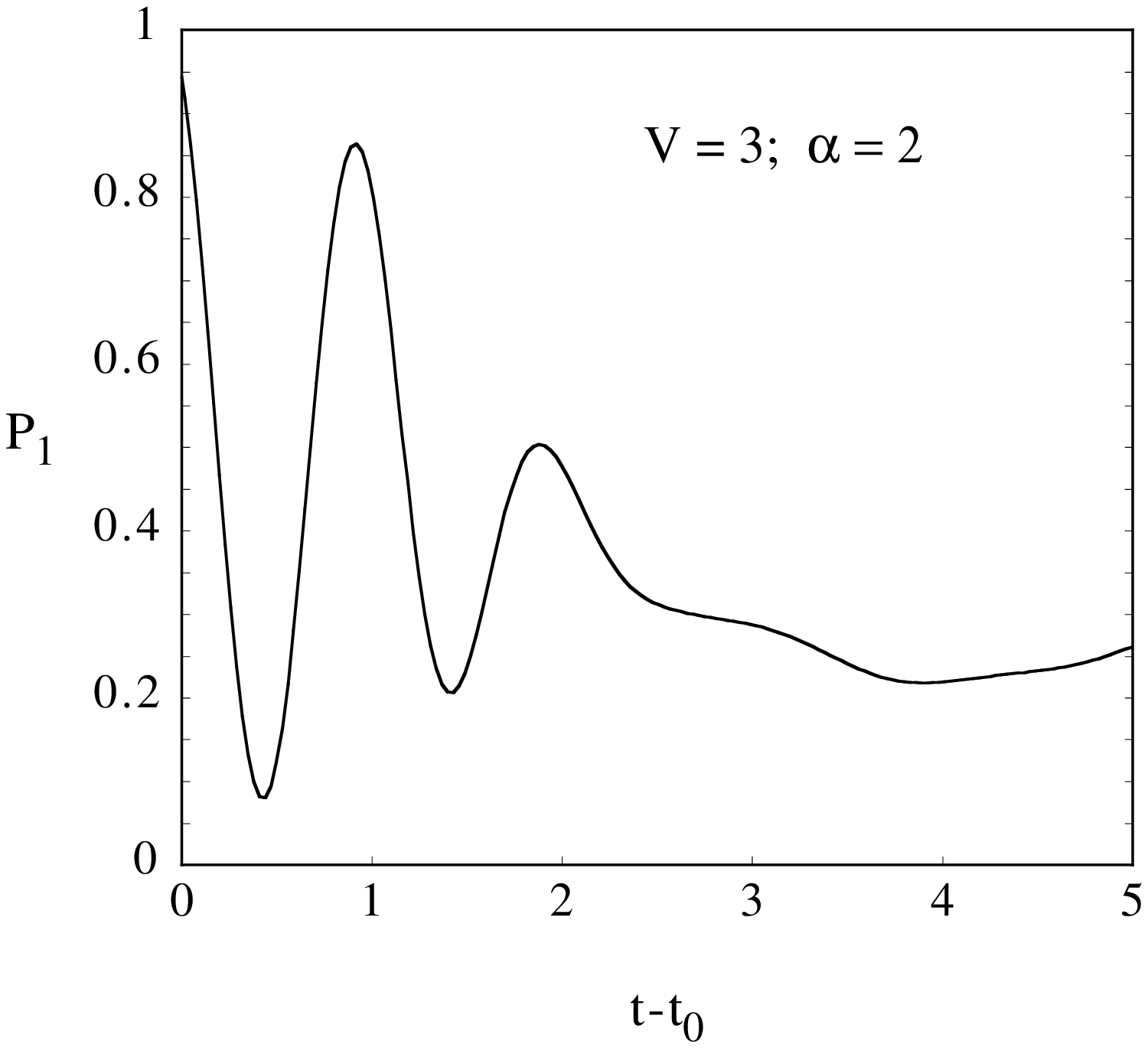}}
\caption[tit]{The occupation $P_1(t)$ for $V_0=3$ and $\alpha=2$. Here
$\Gamma_0\simeq 28$ and the oscillatory behaviour is obvious.}
\label{osc}
\end{figure}

We have especially looked into the possible occurrence of exponential decay and
the emergence of an outgoing wave packet on state $|2\rangle$ in our model. 
Some typical results of our calculations are shown in figs.~\ref{v05}
and~\ref{v07}. They present the logarithm of the occupation $P_1$ of the state
$|1\rangle$ as a function of time. Both figures show descending straight
lines which indicate exponential decay. For very large values of $\Gamma_0$
the initial state occupation $P_1$ becomes clearly oscillatory as demonstrated
in fig.~\ref{osc}. 

\begin{figure}[tbh]
\centerline{\psfig{width=4.0in,file=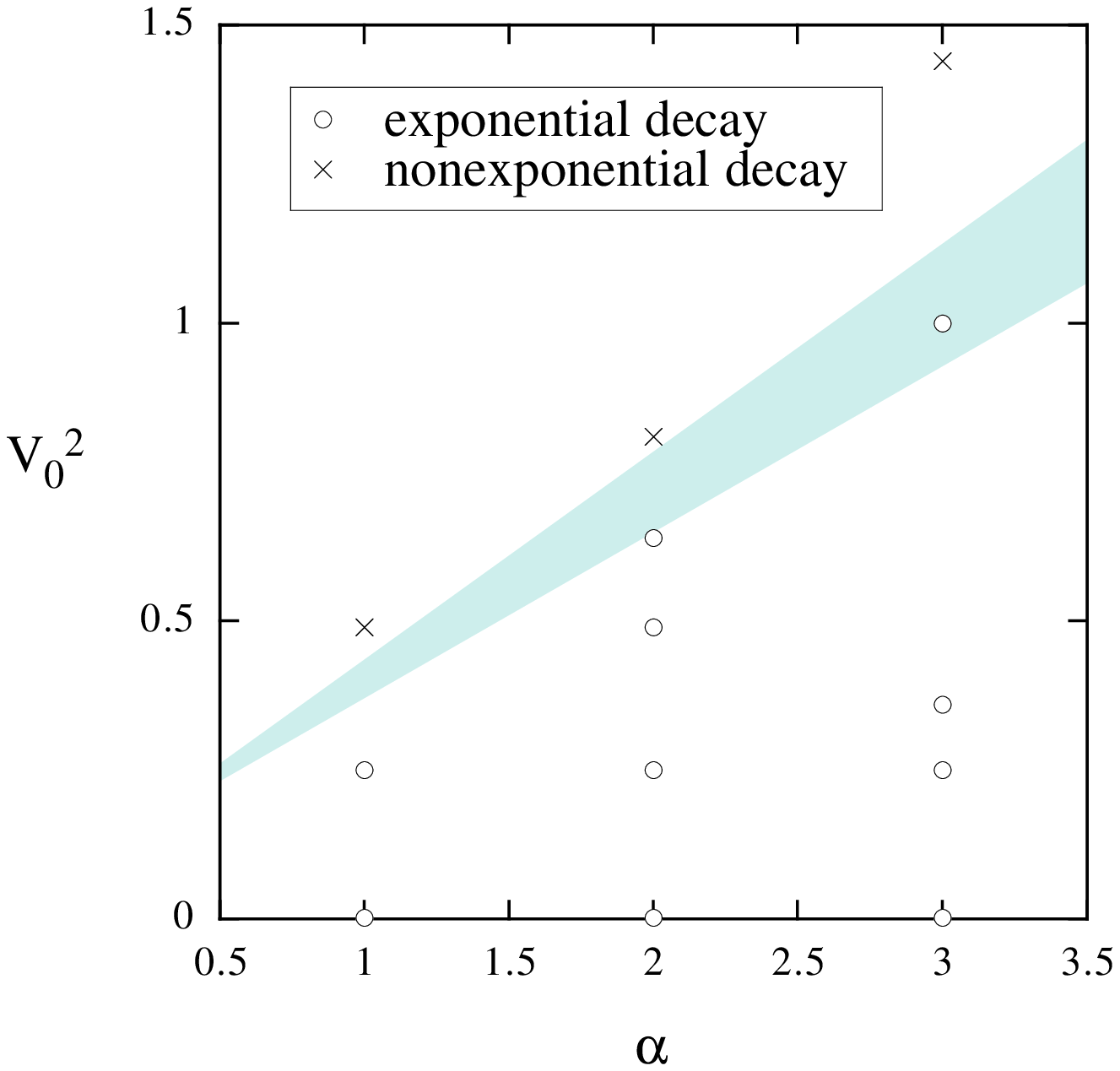}}
\caption[tbt]{The regions of exponential ($\circ$) and nonexponential
($\times$) decay in the ($\alpha$,$V_0^2$) plane. The shaded region
corresponds to the upper limit given in eq.~(\protect\ref{a12}).}
\label{a-v2}
\end{figure}

\begin{figure}[tbh]
\centerline{\psfig{width=5.0in,file=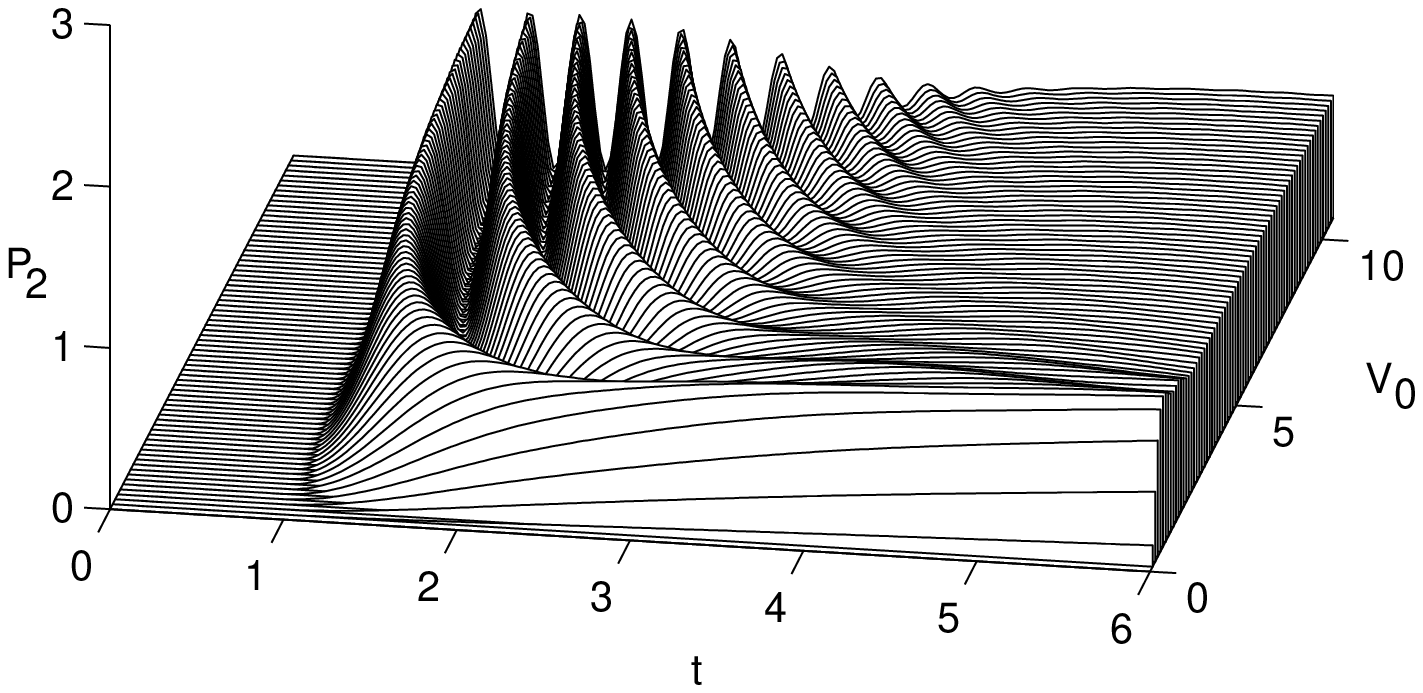}}
\caption[tct]{The time evolution of the excited state occupation $P_2$
for $\alpha=2$ for various values of the coupling $V_0$. The transition
from the steady behaviour to the oscillatory one with increasing $V_0$ is
clear. Here $t_0=1.0$.}
\label{t-v0}
\end{figure}

There is no absolute criterion when exponential decay is observed; the
occurrence of linear behaviour in the semilogarithmic plots is subject to
an arbitrary criterion. Running various parameter combinations, however,
indicates that monotonic exponential decay is observed for values of 
$V_0^2$ smaller than a constant times $\alpha$, thus indicating that the
parameter (\ref{a11}) plays a decisive role in the phenomenon. In
fig.~\ref{a-v2} we show the data points that correspond to exponential and
nonexponential decay in the ($\alpha$,$V_0^2$) plane. The exponential decay
points fall into a region for which we have roughly
\begin{equation}
   \Gamma_0 <2\ldots 2.5,  \label{a12}
\end{equation}
which is thus taken to be the limit of validity of the simple pole
approximation for decay. Typically the nonexponential behaviour appears in
$P_1$ and $P_2$ as oscillations, and the exponential behaviour as a steady
change; this is demonstrated in fig.~\ref{t-v0}.

\begin{figure}[tbh]
\centerline{\psfig{width=5.0in,file=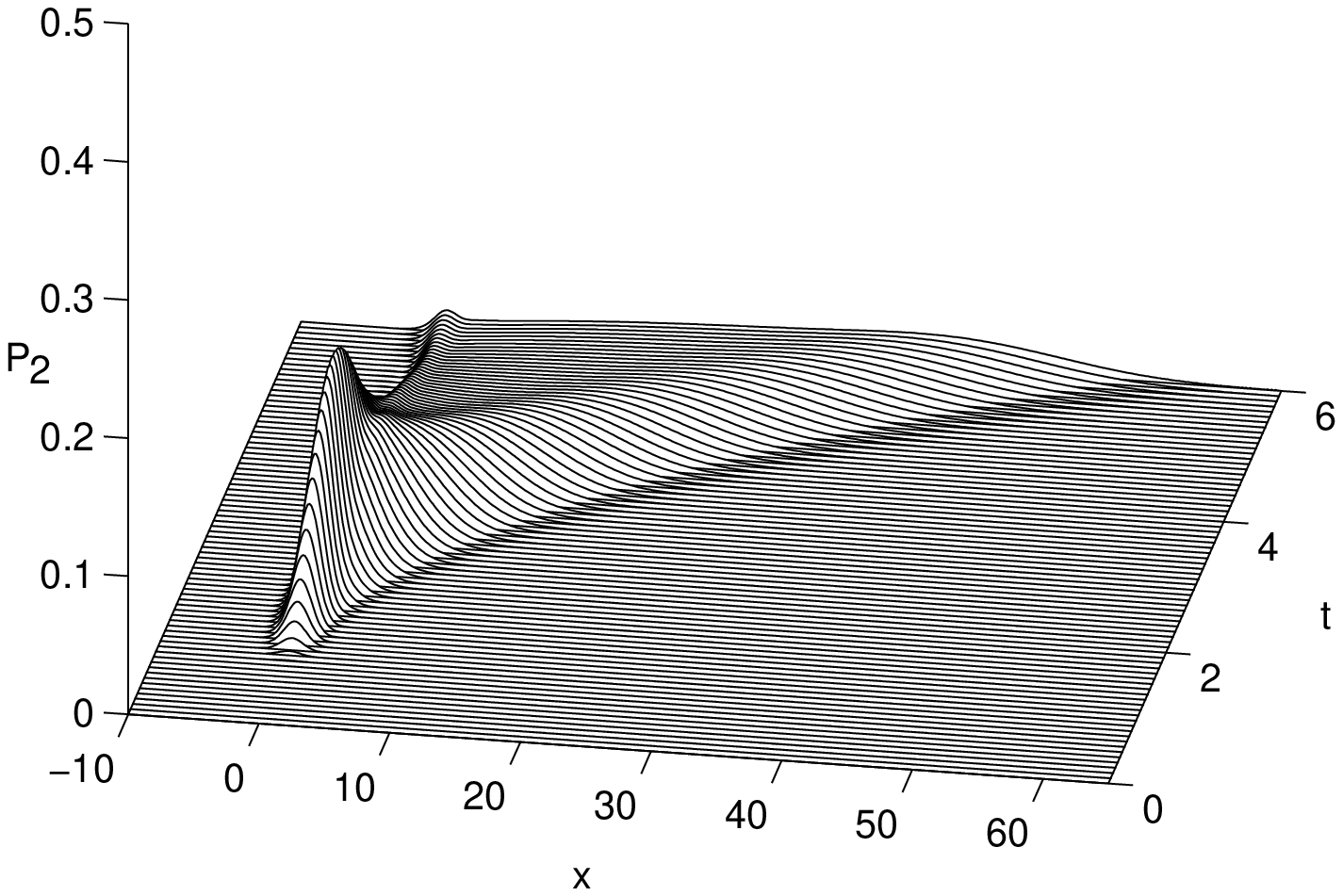}}
\caption[tet]{The excited state wave packet $P_2(x,t)$ for $V=0.7$ and $\alpha
=2$. This is in the exponential decay regime since $\Gamma_0 =1.5$.
Here $t_0=1.0$.}\label{wp07}
\end{figure}

\begin{figure}[tbh]
\centerline{\psfig{width=5.0in,file=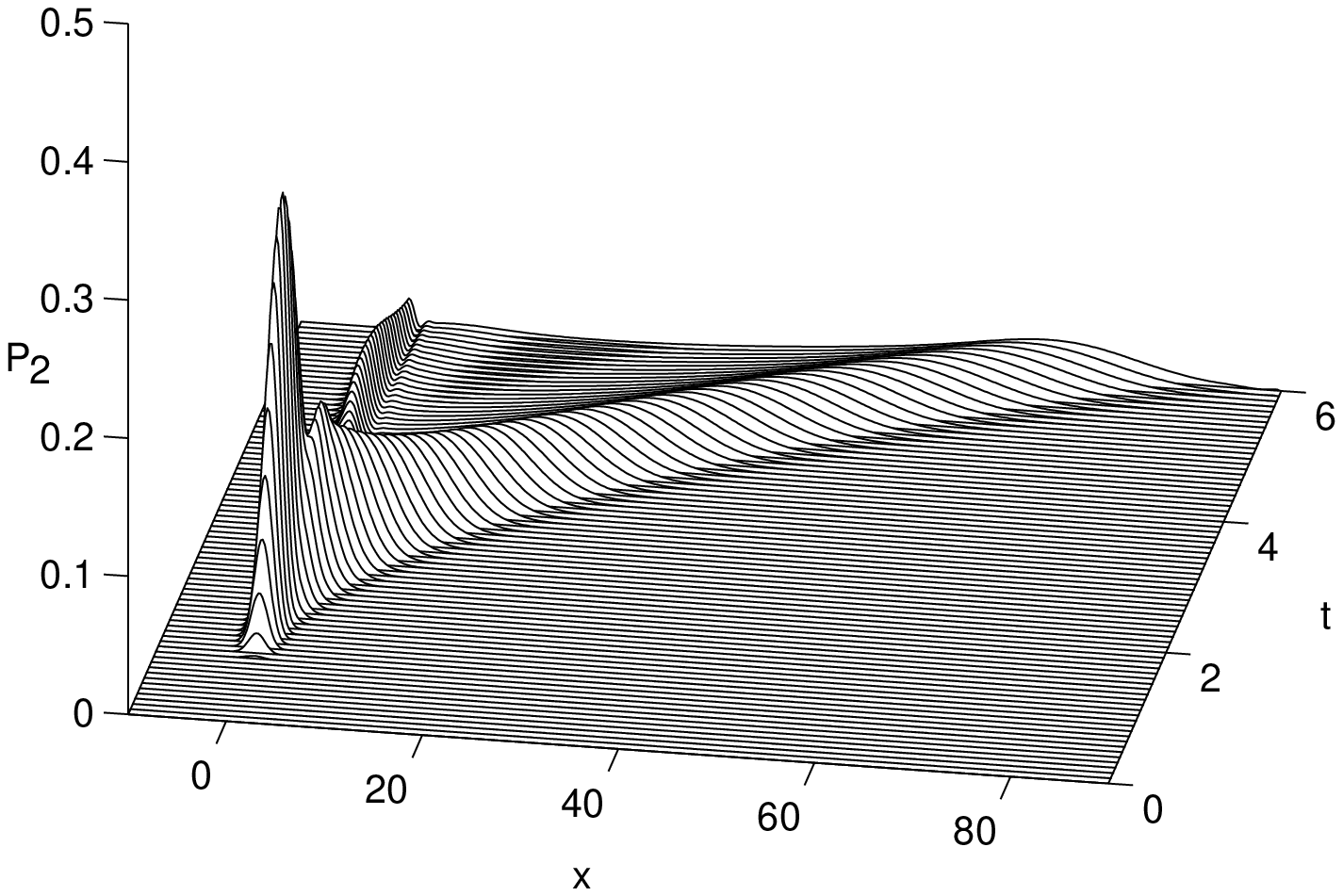}}
\caption[tet]{The excited state wave packet $P_2(x,t)$ for $V_0=1.5$ and $\alpha
=3$. This is in the nonexponential decay regime since $\Gamma_0 =4.7$.}
\label{wp15}
\end{figure}

We have also studied how the emerging probability on level $|2\rangle$
forms a wave packet. For the case $V_0=0.7,\alpha=2$ ($\Gamma_0\simeq 1.5$) we
observe a clear wave crest emerging in fig.~\ref{wp07}. This is followed by a
long, slowly diminishing tail representing the final leakage out of state
$|1\rangle$. We have, however, observed that the formation of a wave packet is
not conditioned on the occurrence of exponential decay. In fig.~\ref{wp15} we
still see a well developed wave packet for the parameters $V_0=1.5$ and
$\alpha=3$ ($\Gamma_0\sim 4.7$) far exceeding that for which perturbation
theory would be expected to work. 

We have not carried out a systematic investigation of the occurrence of a well
formed wave packet. This is as subjective a phenomenon as the exponential
decay. Because we initially find no probability on state $| 2\rangle $ and
ultimately there is no longer any probability leaking out of state $|
1\rangle$, there must be a localized wave packet emerging out of the
system. However, a few words connecting the results of this paper to our
previous work~\cite{SGS92,PSS95,P95,PS97} are appropriate here. 

In both figures~\ref{wp07} and~\ref{wp15}
we see a clear, peaked wave crest wave packet. On the other hand,
figures~\ref{v05} and~\ref{v07} show that for all parameter values the ground
state occupation behaviour is not exponential when $t\simeq t_0$. This is
because initially the excitation process follows the well known area
theorem, which for two resonant levels and a coupling $V_0$ states that
$P_1=1-P_2=\cos^2[V_0(t-t_0)]\simeq 1-V_0^2(t-t_0)^2+{\cal O}[(t-t_0)^4]$ for
$t\geq t_0$. This is the origin of Rabi oscillations (if the levels are
off resonance, then the oscillation frequency and amplitude are modified).
For more details see e.g.~\cite{r4}.
These oscillations simply indicate that both states are ``equal", i.e.,
transitions can take place also back from state $|2\rangle$ to state
$|1\rangle$. In the Weisskopf-Wigner regime the second state acts like a 
reservoir and population can move only from state $|1\rangle$ to state
$|2\rangle$. 

This is the point where the dynamics of the wave packet on state $|2\rangle$ 
comes to the rescue. In our earlier studies we have given special attention to 
cases where the dynamics inhibits the Rabi oscillations, because the excited 
state population is accelerated away from the resonance region faster than the 
Rabi oscillations take place~\cite{SGS92,PSS95,P95,PS97}. Although our earlier 
studies were made for short pulses, the basic principle is the same. Thus the 
reservoir model works best for the excitation when $\alpha$ is large and wave 
packet motion on state $|2\rangle$ takes place swiftly after excitation.

We have now established a loose criterion for obtaining exponential decay. Next
we need to consider the validity of the perturbation result~(\ref{a10}). The
exponential behaviour is only a necessary but not a sufficient condition for
eq.~(\ref{a10}) to hold. For suitably large values of $V_0$ the perturbation
result may fail even if the criterion~(\ref{a12}) is fulfilled and the decay
is exponential. For this purpose we calculate the full perturbation result for
our model in the following section.

\section{Analytic considerations}\label{sec:analytic}

The perturbative rate~(\ref{a10}) contains the Condon factor 
\begin{equation}
   S_{01}=\langle \varphi_0| \psi_{E_0}\rangle ,  \label{a13}
\end{equation}
which is the overlap of the initial and final states; see e.g.
refs.~\cite{Condon,Coolidge,r8,Dowling}. This scalar product is
most efficiently calculated in momentum space, where we need the Fourier
transform of the state~(\ref{a4}). Calculating the continuum wave function
$\psi_{E_0}$ for the energy~(\ref{a3}) we set ${\cal E}=0$, which gives 
\begin{eqnarray}
   S_{01} & = & \frac{\textstyle 2^{5/8}\pi^{1/4}}{\textstyle\sqrt{\alpha
      }}\int\frac{\textstyle dk}{\textstyle 2\pi} 
      \exp \left[ -\left( \frac{k^2}{\sqrt{2}}+i\frac{k^3}{3\alpha }\right)
      \right]\nonumber\\ 
      &  & \nonumber \\ 
      & = & \frac{e^{\alpha ^2/3\sqrt{2}}}{2^{3/8}\pi^{3/4}\sqrt{\alpha }} 
      \int_{-\infty }^\infty du\,\exp \left[ i\left( \frac{\alpha
      u}2+\frac{u^3}{ 
      3\alpha }\right) \right] ,
   \label{a14}
\end{eqnarray}
where we have used the substitution $k=i\alpha/\sqrt{2}-u$. Using the Airy 
function 
\begin{equation}
   {\rm Ai}(x)=\int_{-\infty }^\infty \frac{ds}{2\pi}\exp \left[ i\left(
   \frac{s^3}{3} + sx\right) \right] ,  \label{a15}
\end{equation}
we can write (after setting $u=\alpha^{1/3}s$ in eq.~(\ref{a14}))
\begin{equation}
   S_{01}=\frac{2^{5/8}\pi ^{1/4}e^{\alpha ^2/3\sqrt{2}}}{\alpha ^{1/6}} 
   {\rm Ai}\left( \frac{\alpha ^{4/3}}2\right).  \label{a16}
\end{equation}
The result~(\ref{a16}) seems to contradict the simple dependence on $\alpha$
from~(\ref{a14}), but the asymptotic relation ${\rm Ai}(x)\simeq x^{-1/4}\exp
(-2x^{3/2}/3)/\sqrt{4\pi}$, valid for large $x$, restores the correct scaling
with $\alpha$ in the perturbative limit. In fact, the asymptotic result
for the Franck-Condon factor becomes
\begin{equation}
   \lim_{\alpha\rightarrow\infty} |S_{01}|^2 = 
   \frac{1}{\alpha(2\pi^2)^{1/4}}.
\end{equation}
This is exactly what the Condon reflection principle would 
predict~\cite{Condon,r8,Burnett}. According to it 
\begin{equation}
   \lim_{\alpha\rightarrow\infty} |S_{01}|^2 = \left|\frac{d(U_2-U_1)}
   {dx}\right|_{x=x_0}^{-1}|\varphi_0(x=x_0)|^2,
\end{equation}
where $x_0$ is the point where $U_2(x_0)=U_1(x_0)$; in our case $x_0$
approaches zero as $\alpha\rightarrow\infty$, and our ground state
wave function contribution becomes $|\varphi_0(x=0)|^2=(2\pi^2)^{-1/4}$. 
The reflection
principle simply means that the excited state wave functions for steep
potentials are so localised to $x=x_0$ that they behave like $\delta$ functions
inside the overlap integrals. Thus we get that
\begin{equation}
   \lim_{\alpha\rightarrow\infty} \Gamma = \frac{1}{(2\pi^2)^{1/4}}\Gamma_0.
   \label{Gpert}
\end{equation}

\begin{figure}[tbh]
\centerline{\psfig{width=4.0in,file=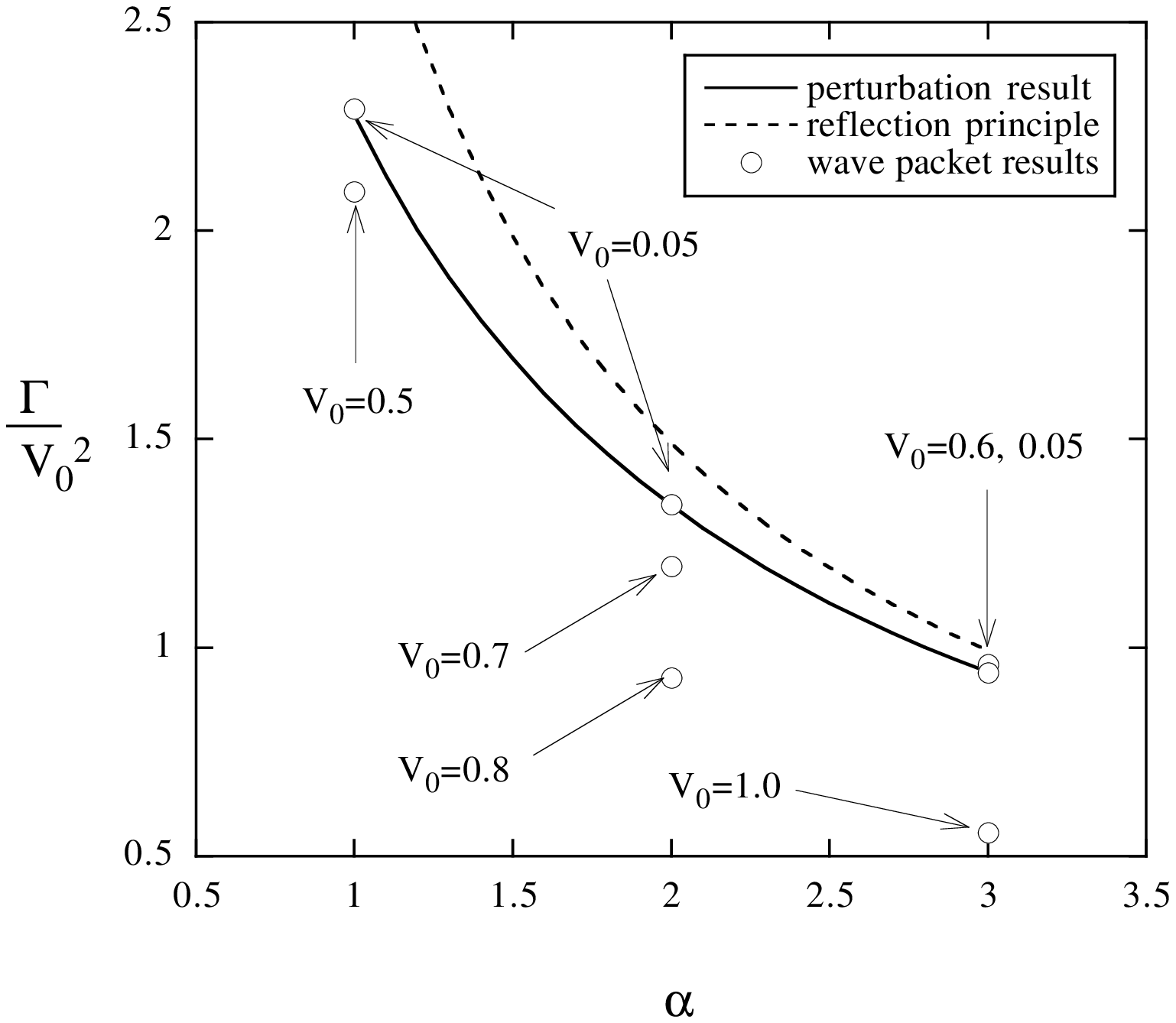}}
\caption[tyt]{The $\alpha$ dependence of the decay rate $\Gamma$. We have
plotted the wave packet results (open circles), the full perturbation 
result (solid line) [eqs.~(\protect\ref{a10}) and~(\protect\ref{a16})] 
and the asymptotic result (dashed line) [eq.~(\protect\ref{Gpert})]. 
All selected data points correspond to exponential decay.}\label{kal1}
\end{figure}

The interesting conclusion from eq.~(\ref{a14}) is that the decay rate is not
the simple function of $\alpha$ suggested by eq.~(\ref{a11}) except in the
asymptotic limit. The behaviour of $\Gamma(\alpha)$, however, is found to be 
simply monotonic with increasing $\alpha$ as shown in the fig.~\ref{kal1}.

\section{Discussion}\label{sec:concl}

We have shown that in the perturbation limit and for steep excited state 
potentials the Weisskopf-Wigner model can describe adequately laser induced 
molecular excitation processes. Our numerical results are in good agreement
with analytical perturbation theory results in the parameter region where
we expect them to be applicable. 

Although we have considered mainly the molecular excitation process, there are
some interesting atomic systems that resemble the molecular model shown in 
fig.~\ref{schemes}(b). One can trap single ions electromagnetically, and cool 
them down to the lowest motional state of the trap potential, which in this 
energy range corresponds to a harmonic
potential~\cite{StigJMO,Monroe,Zoller,Knight}.
Furthermore, one can also trap neutral atoms with magnetic fields, and cool
them evaporatively into the density-temperature region where Bose-Einstein
condensation takes place~\cite{Anderson,Ketterle1}. The behaviour of the 
condensed atoms can be described as a single wave 
function~\cite{Keith,Mark}. Our model, therefore, resembles the
recent studies for output couplers for the condensates, in which rf fields
couple the trapping state to a nontrapping (continuum) state---such a system 
can be regarded as a first generation atom laser~\cite{CouplerPRL,Alaser}. 
Thus our model may provide some insights when one wishes to move from 
pulsed atom lasers towards cw atom lasers.

\section*{Acknowledgments}

The authors thank the Academy of Finland for financial support, and the
Finnish Centre for Scientific Computing for providing the computer facilities.

\setlength{\baselineskip}{4mm}
\renewcommand{\baselinestretch}{0.8} 
\setlength{\itemsep}{0.0mm}
\setlength{\parsep}{2.0mm}

\end{document}